\documentstyle[twocolumn,eqsecnum,aps]{revtex}
\def\be{\begin{eqnarray}}
\def\ee{\end{eqnarray}}
\def\bea{\begin{eqnarray}}
\def\eea{\end{eqnarray}}
\def\bp{{\bf b_\perp}}

\def\Dp{{\bf \Delta_\perp}}
\def\Dps{{\bf \Delta^2_\perp}}
\begin{document}
\draft
\title{Impact Parameter Dependent Parton Distributions and
Transverse Single Spin Asymmetries}
\author{Matthias Burkardt}
\address{Department of Physics\\
New Mexico State University\\
Las Cruces, NM 88003-0001\\U.S.A.}
\maketitle
\begin{abstract}
Generalized parton distributions (GPDs) with purely
transverse momentum transfer can be interpreted as
Fourier transforms of the distribution of 
partons in impact parameter space. The helicity-flip
GPD $E(x,0,-\Dps)$ is related to the distortion
of parton distribution functions in impact parameter
space if the target is not a helicity eigenstate,
but has some transverse polarization. This transverse
distortion can be used to develop an intuitive
explanation for various transverse single spin
asymmetries. 
\end{abstract}
\narrowtext
\section{Introduction}
Deep-inelastic scattering experiments allow
the determination of parton distribution functions 
(PDFs), which have the very physical interpretation
as momentum (fraction) distributions in the
infinite momentum frame (IMF).
PDFs are defined as the forward matrix element of 
a light-like correlation function, i.e.
\be 
q(x) &=&\left\langle P,S\left| 
\hat{O}_q(x,{\bf 0}_\perp) 
\right|P,S\right\rangle
\label{eq:pd}\\ 
\Delta q(x)S^+ &=& P^+
\left\langle P,S\left|\hat{O}_{q,5}
(x,{\bf 0}_\perp) 
\right|P,S\right\rangle 
\nonumber
\ee
with
\be
& &\hat{O}_q(x,{\bf 0}_\perp) 
\equiv \int\!\! \frac{dx^-}{4\pi}
\bar{q}(-\frac{x^-}{2},{\bf 0}_\perp)
\gamma^+ q(\frac{x^-}{2},{\bf 0}_\perp)
e^{ix{p}^+x^-}\label{eq:0perp}\\
& &\hat{O}_{q,5}(x,{\bf 0}_\perp) 
\equiv \int\!\! \frac{dx^-}{4\pi}
\bar{q}(-\frac{x^-}{2},{\bf 0}_\perp)
\gamma^+ \gamma_5
q(\frac{x^-}{2},{\bf 0}_\perp)
e^{ix{p}^+x^-} .\nonumber
\ee
When sandwiched between states that have the same
light-cone momentum $p^+=\frac{1}{\sqrt{2}}(
p^0+p^3)$, these operators act as a `filter' for
quarks of flavor $q$ with momentum fraction $x$.
Throughout this work, we will use light-cone gauge 
$A^+=0$. In all other gauges, a straight line gauge 
string connecting the quark field operators needs to
be included in this definition (\ref{eq:pd}).
Obviously, since PDFs are expectation values
taken in plane wave states, they
contain no information about the position space
distribution of quarks in the target.

Generalized parton distributions (GPDs)\cite{GPD}, 
which describe for example the scaling limit in 
real and virtual Compton scattering experiments,
are defined very similar to PDFs except that one 
now takes a non-forward matrix element of the
light-cone correlator
\be
& &\langle P^\prime,S^\prime|
\hat{O}_q(x,{\bf 0}_\perp) 
|P,S\rangle \label{eq:gpd}\\
& &= \frac{1}{2\bar{p}^+}
\bar{u}(p^\prime,s^\prime)\left(\gamma^+  
H_q(x,\xi,t)
+ i\frac{\sigma^{+\nu}\Delta_\nu}{2M} E_q(x,\xi,t)
\right)u(p,s) 
\nonumber\\[1.ex]
& &
\langle P^\prime,S^\prime|
\hat{O}_{q,5}(x,{\bf 0}_\perp) 
|P,S\rangle \label{eq:gpd2}\\
& &
= \frac{1}{2\bar{p}^+}
\bar{u}(p^\prime,s^\prime)\left(\gamma^+\gamma_5  
\tilde{H}_q(x,\xi,t)
+ i\frac{\gamma_5\Delta^+}{2M}\tilde{E}(x,\xi,t)
\right)u(p,s)
\nonumber
\ee
with $\bar{p}^\mu = \frac{1}{2}\left( p^\mu
+p^{\prime \mu}\right)$ being the mean momentum
of the target,
$ \Delta^\mu = p^{\prime \mu}-p^\mu$ the four 
momentum transfer, and $t=\Delta^2$ the invariant
momentum transfer. The skewedness parameter 
$\xi = -\frac{\Delta^+}{2\bar{p}^+}$ quantifies 
the change in light-cone momentum.
 
 
An important physical interpretation for
GPDs derives from the fact that they are the form 
factors of the light-cone correlators 
$\hat{O}_q(x,{\bf 0}_\perp)$ and
$\hat{O}_{q,5}(x,{\bf 0}_\perp)$.
Because of that, and by
analogy with ordinary form factors, one would
therefore expect that GPDs can be interpreted as
some kind of Fourier transform of
parton distributions in position space. Indeed,
as has been shown in Ref. \cite{me1,me,diehl2}, the
helicity non-flip\footnote{The `helicity' basis that
we are using refers to the infinite momentum frame
helicity \cite{soper}.}
GPD $H$ for $\xi=0$ is the
Fourier transform of the (unpolarized) impact 
parameter dependent parton distribution function
$q(x,\bp)$, i.e.
\be
q(x,\bp) = \int \frac{d^2 \Dp}{(2\pi)^2}
e^{-i\Dp \cdot \bp} H(x,0,-\Dps).
\label{eq:q}
\ee
The reference point for the impact parameter
in Eq. (\ref{eq:q}) 
is the (transverse) center of momentum (CM)
of the target
\be
{\bf R_\perp} \equiv \frac{1}{p^+}
\int d^2{\bf x_\perp} \int dx^- T^{++} {\bf x_\perp}
= \sum_{i\in q,g} x_i {\bf r_{\perp,i}},
\label{eq:Rperp}
\ee
where $T^{++}$ is the light-cone momentum
density component of the energy
momentum tensor. The sum in the parton representation for
${\bf R_\perp}$ extends over the transverse
positions ${\bf r_{\perp,i}}$ of all quarks
quarks and gluons in the target and the weight
factors $x_i$ is the momentum fraction carried by 
each parton.
The impact parameter dependent PDFs are defined
by introducing the $\bp$-dependent light-cone
correlation
\be
& &\!\!\!\!\!\!\!\!q(x,\bp) \equiv
\\
& &\quad \quad\quad \left\langle p^+,
{\bf R_\perp}= {\bf 0_\perp},\lambda \left|
\hat{O}_q(x,\bp )\right|p^+,{\bf R_\perp}
= {\bf 0_\perp},\lambda\right\rangle,
\!\!\!\!\!\!\!\!\!
\nonumber
\ee
where
\be
\left|p^+,{\bf R_\perp}
= {\bf 0_\perp},\lambda\right\rangle
\equiv {\cal N} \int d^2{\bf p_\perp}
\left|p^+,{\bf p_\perp},\lambda\right\rangle
\label{eq:local}
\ee
is a state whose transverse CM is localized at the 
origin and ${\cal N}$ is a normalization constant.
They are simultaneous eigenstates of the
light-cone momentum $p^+$, the transverse CM 
(with eigenvalue ${\bf 0_\perp}$ and
the angular momentum operator $J_z$, which is 
possible due to the Galilean subgroup of transverse
boosts in the IMF \cite{soper}.

A similar connection exists between $\tilde{H}$
and impact parameter dependent polarized
PDFs
\be
\Delta q(x,\bp) = \int \frac{d^2 \Dp}{(2\pi)^2}
e^{-i\Dp \cdot \bp} \tilde{H}(x,0,-\Dps),
\label{eq:Dq}
\ee
where
\be
& &\!\!\!\!\Delta q(x,\bp)\equiv\\
& & 
\left\langle p^+,{\bf R_\perp}
= {\bf 0_\perp},\uparrow \left|
\hat{O}_{q,5}(x,\bp )\right|p^+,{\bf R_\perp}
= {\bf 0_\perp},\uparrow\right\rangle.
\!\!\!\!\!\!\!\!\!\!\!\!\!\!\!\!\!\!\!\!
\nonumber
\ee
It should be emphasized that impact parameter
dependent parton distributions have an 
interpretation as a probability density. In fact
\be
\int d^2\bp q(x,\bp)&=&q(x)\nonumber\\
q(x,\bp) &\geq& 0 \quad \quad \quad(x>0)\nonumber\\
q(x,\bp) &\leq& 0 \quad \quad \quad(x<0)\nonumber\\
\int d^2\bp \Delta q(x,\bp)&=&\Delta q(x) \nonumber\\
\left|  \Delta q(x,\bp) \right|
&\leq&  \left|q(x,\bp)\right| .
\ee
Eqs. (\ref{eq:q}) and (\ref{eq:Dq}) imply that
GPDs for $\xi=0$ can be used to construct
`tomographic images' \cite{ralston} of
the target nucleon, where one can study
`slices' of the nucleon in impact parameter 
space for different values of the light-cone
momentum fraction $x$, and one can 
learn how the size of the nucleon depends
on $x$. Another useful piece of information
that is contained in these 3-dimensional
images is how the light-cone momentum 
distribution of the quarks
varies with the distance from the CM.

Amazingly, the transverse resolution in these
images is not limited by relativistic effects,
but only by the inverse momentum of the photon
that is used to probe the GPDs, which
determines the pixel size in these images.

\section{GPDs with Helicity Flip}
\label{sec:gpdflip}
In order to develop a probabilistic 
interpretation for $E(x,0,t)$, it is
necessary to consider helicity flip amplitudes
because otherwise $E(x,\xi=0,t)$ 
does not contribute \cite{flip}
\footnote{The helicity labels 
$\uparrow$, $\downarrow$ in Eqs. 
(\ref{eq:noflip}) and (\ref{eq:flip}) refer to
helicity states in the IMF \cite{soper}.}
\bea
\label{eq:noflip}\\
\left\langle p+,{\bf p_\perp}+\Dp, \uparrow \left| 
\hat{O}_q(x,{\bf 0_\perp})
\right| p^+,{\bf p_\perp},\uparrow\right\rangle 
&=& H(x,0,-{\bf \Delta}_\perp^2) .
\nonumber\\
& &\nonumber\\
\label{eq:flip}
\left\langle p^+,{\bf p_\perp}+\Dp, \uparrow
\left| \hat{O}_{q,5}(x,{\bf 0_\perp})
\right|p^+,{\bf p_\perp},\downarrow\right\rangle
\!\!\!\!& & \\
& &\!\!\!\!\!\!\!\!\!\!\!\!\!\!\!\!
\!\!\!\!\!\!\!\!\!\!\!\!\!\!\!\!\!\!\!\!
\!\!\!\!\!\!\!\!\!\!\!
= -\frac{\Delta_x-i\Delta_y}{2M}E(x,0,-{\bf \Delta}_\perp^2) .
\nonumber
\eea
Therefore, if one wants to develop a
density interpretation for $E(x,0,-\Dps)$ one
needs to consider states that are not helicity
eigenstates.
The superposition where the contribution from
$E$ is maximal corresponds to states
where $\uparrow$ and $\downarrow$ contribute
with equal magnitude. We thus consider the state
\be
\left| X \right\rangle \equiv
\frac{1}{\sqrt{2}}\left[ 
\left|p^+,{\bf R_\perp}={\bf 0_\perp},\uparrow
\right\rangle +
\left|p^+,{\bf R_\perp}={\bf 0_\perp},\downarrow
\right\rangle\right] ,
\ee
which one may interpret as a state that is
`polarized in the $x$ direction (in the IMF)'.
However, since the notion of a transverse 
polarization is somewhat tricky in the basis
that we are using (states that are eigenstates of
$p^+$ and ${\bf R_\perp}$), there may be some
relativistic corrections to the actual 
interpretation of what this state corresponds to.
In the following, we will keep this caveat in mind
when studying the properties of this state even
though we will refer to this state as a
'transversely polarized nucleon (in the IMF)'.
The unpolarized impact parameter dependent PDF in
this state will be denoted $q_X(x,\bp)$.

Repeating the same steps that led to Eq. (\ref{eq:q})
and using Eqs. (\ref{eq:noflip}) and (\ref{eq:flip}),
one finds
\bea
q_X(x,\bp)&\equiv&\left\langle X \right|
\hat{O}_q(x,\bp)\left|X\right\rangle
\nonumber\\
& & \!\!\!\!\!\!\!\!\!\!\!\!\!\!\!\!\!\!\!\!
\!\!\!\!\!\!\!\!\!\!\!\!\!\!\!=\!
\int\!\!\frac{d^2\Dp}{(2\pi)^2}
e^{-i\Dp\cdot\bp}
\left[ H_q(x,0,\!-\Dps) 
+ \frac{i\Delta_y}{2M} E_q(x,0,\!-\Dps)\right]
\nonumber\\
&=& q(x,\bp) - \frac{1}{2M}\frac{\partial}
{\partial b_y} {\cal E}_q(x,\bp),
\label{eq:qX}
\eea
where we denoted ${\cal E}_q$ the Fourier transform
of $E_q$, i.e.
\be
{\cal E}_q(x,\bp)\equiv
\int \frac{d^2\Dp}{(2\pi)^2}e^{-i\Dp\cdot\bp}
E_q(x,0,-\Dps).
\ee
Physically, what this result means is that for a
nucleon that is transversely polarized and moves
with a large momentum,
an observer at rest sees parton distributions 
that are distorted sideways in the transverse plane.
Obviously, for transversely polarized nucleons the
axial symmetry of the problem is broken and the
impact parameter dependent PDFs no longer need to
be axially symmetric.
The direction of the distortion is perpendicular
to both the spin and the momentum of the nucleon.
\footnote{Note that ${\vec S}\times {\vec p}$ 
transforms like a position space vector ${\vec r}$
under $P$ and $T$ transformations.}
Although the distortion is mathematically described by
Eq. (\ref{eq:qX}) in a model-independent way, it is 
instructive to consider a semi-classical picture
for the effect where the physical origin of this 
distortion results from a superposition
of translatory and orbital motion of the partons when the
nucleon is polarized perpendicular to its direction 
of motion.
If the spin of the nucleon is ``up'' (looking into the 
direction of motion of the nucleon) and the orbital angular
momentum of the quarks is parallel to the nucleon spin 
then the orbital motion adds to the momentum on the
right side of the nucleon and subtracts on the left side, 
i.e. partons on the right side get boosted to larger momentum 
fractions $x$ and on the left they get decelerated to smaller 
$x$ (compared to longitudinally polarized nucleons).
Since parton distributions decrease with $x$ (at large momenta
they drop like a power of $x$ and at small $x$ they grow like an
inverse power of $x$), boosting all partons on one side of the 
nucleon results in an increase of the number of partons at a fixed
value of $x$ on that side, while the opposite effect occurs on the
other side.
Therefore, the acceleration/deceleration due to the superposition of
the orbital with the translatory motion
results in an increase of partons
on the right and a decrease on the left, i.e. the net result is that
the parton  distribution in the transverse plane has been shifted or 
distorted to the right.
Of course, for quarks with orbital angular momentum antiparallel
to the nucleon spin the direction of the distortion is reversed 
(to the left). In Ref. \cite{ji} it has been shown that the helicity 
flip GPD $E$ is related to the angular momentum carried by the quark. 
This result, together with the above semiclassical description 
about the physical origin of the distortion, provides an intuitive 
explanation for the fact that this distortion is described by $E$.

It should be emphasized that transverse asymmetries in
impact parameter dependent PDFs are consistent with time-reversal
invariance since ${\vec b}\cdot ({\vec p}\times {\vec S})$ is 
invariant under $T$.
In contrast, ${\vec k}\cdot ({\vec p}\times {\vec S})$ is {\it not} 
invariant under $T$, and therefore transverse asymmetries in
unintegrated parton densities $q(x,{\bf k}_\perp)$ are only
permitted if final state interaction effects are incorporated
into the definition of unintegrated parton densities \cite{collins2}.

Unfortunately, little is known about generalized parton distributions
and it is therefore in general difficult to make predictions without
making model assumptions. However, it {\it is} possible to make
a model independent statement about the resulting transverse flavor
dipole moment
\be
d^y_q &\equiv& \int dx \int d^2 {\bf b}_\perp
q_X(x, \bp) b_y \nonumber\\
&=& -\frac{1}{2M} \int dx \int d^2 {\bf b}_\perp
b_y \frac{\partial}{\partial b_y} {\cal E}_q(x,\bp)
\nonumber\\
&=& \frac{1}{2M} \int dx \int d^2 {\bf b}_\perp {\cal E}_q(x,\bp)
= \frac{1}{2M} \int dx E_q(x,0,0) \nonumber\\
&=& \frac{F_{2,q}(0)}{2M},
\label{eq:dy}
\ee
where we used that the integral of $E_q$ yields the
Pauli formfactor $F_{2,q}$ for flavor $q$ \cite{ji}. For $u$ and $d$ quarks,
$F_{2,q}(0)\equiv \kappa_{q/p}$ in the proton is of the order of
$\left|\kappa_{q/p}\right|\sim 1-2$ (for a more detailed estimate
see Appendix \ref{app}), i.e. the resulting transverse flavor
dipole moments are on the order of
\be
d^y_q \sim 0.1-0.2 \,\mbox{fm}.
\ee
In fact, using only isospin symmetry, one finds for a transversely
polarized proton (\ref{eq:kappaud})
\be
d^y_u-d^y_d = \frac{\kappa_{u/p} - \kappa_{d/p}}{2M}
\approx 0.4  \,\mbox{fm},
\ee
i.e. the flavor center for $u$ and $d$ quarks  get separated in
opposite directions to the point where the separation is of the
same order as the expected size of the valence quark distribution.
\footnote{It should be emphasized that the transverse center
of momentum of the whole nucleon does not shift since 
$\sum_{i\in q,g} \int dx x E_i(x,0,0)=0$ if one sums
over the contributions from all flavors as well as from the glue
\cite{anom}.}

In order to illustrate the magnitude of the distortion graphically, 
we make a simple model for the $\Dp$ dependence of GPDs \cite{me}
\be
H_q(x,0,-{\bf \Delta}_\perp^2) = q(x)
e^{-a{\bf \Delta}_\perp^2(1-x)\ln \frac{1}{x}} .
\label{eq:log2}
\ee
This ansatz incorporates both the expected large $x$ behavior
($H_q$ should become $x$-independent as $x\rightarrow 1$) and
the small $x$ behavior (Regge behavior). Furthermore, in the
forward limit ($\Dp=0$), $H_q$ reduces to the unpolarized PDF
$q(x)$. In impact parameter space this ansatz implies
\be
q(x,{\bf b_\perp^2}) = q(x)\frac{1}{4\pi a (1-x)\ln \frac{1}{x}}
e^{-\frac{ {\bf b}_\perp^2 }{4a(1-x)\ln \frac{1}{x}}} .
\label{eq:model}
\ee
\begin{figure}
\unitlength1.cm
\begin{picture}(10,13.5)(3.7,2.7)
\includegraphics{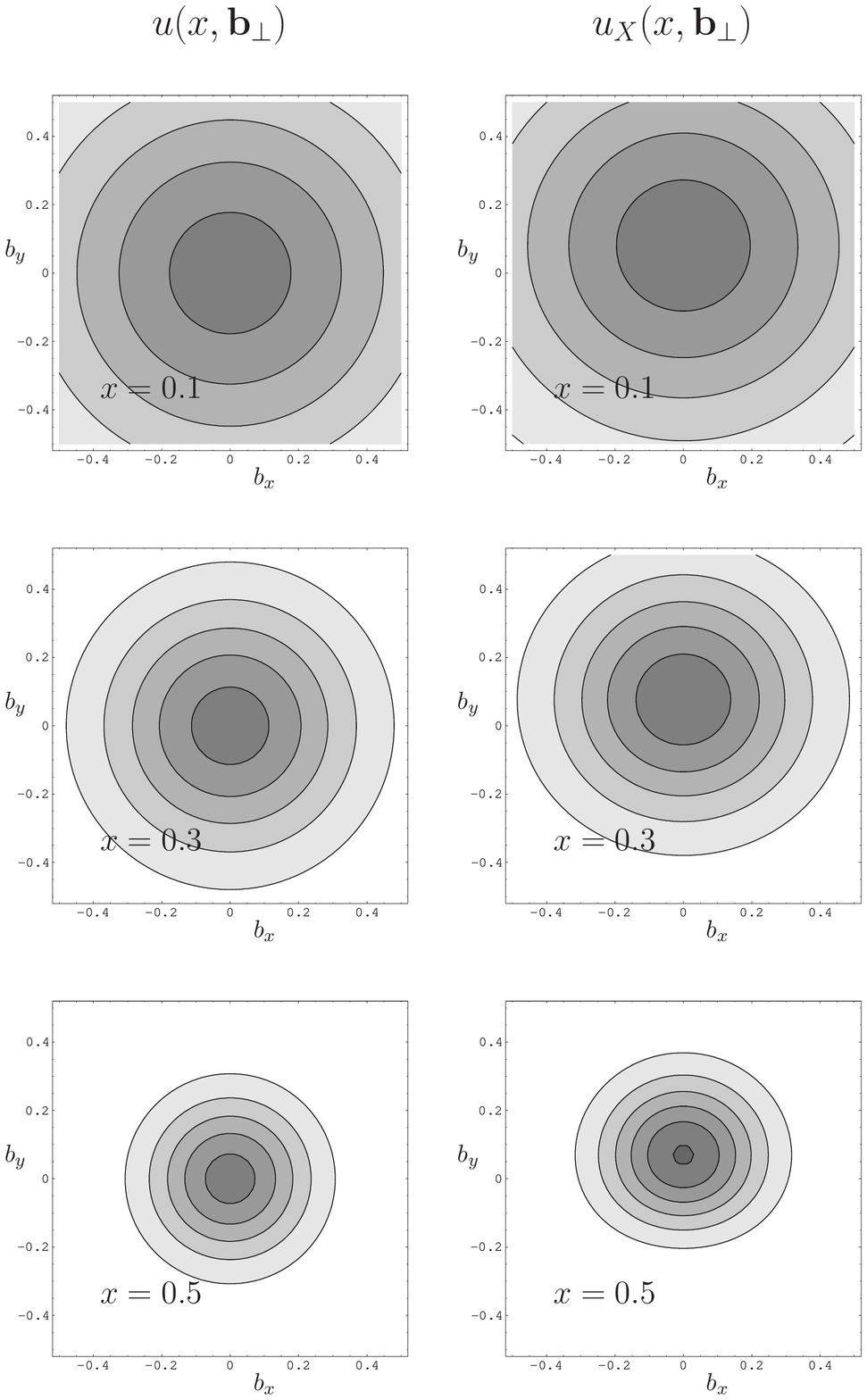}
\end{picture}
\caption{$u$ quark distribution in the transverse
plane for $x=0.1$, $0.3$, and $0.5$ (\ref{eq:model}).
Left column: $u(x,{\bf b_\perp})$, i.e. the
$u$ quark distribution for unpolarized 
protons; right column: $u_X(x,{\bf b_\perp})$, i.e.
the unpolarized $u$ quark distribution for
`transversely polarized'
protons $\left|X\right\rangle = 
\left|\uparrow\right\rangle +
\left|\downarrow\right\rangle$.
The distributions are normalized to the central 
(undistorted) value $u(x,{\bf 0_\perp})$.}
\label{fig:panelu}
\end{figure}  
For the helicity flip distributions $E_q$ we assume that the
$\Dp$ dependence is the same as for $H_q$ and we fix the overall 
normalization by demanding that the integral of $E_q(x,0,0)$ yields
the anomalous magnetic moments
\bea
E_u(x,0,t)&=&\frac{1}{2}\kappa_u H_u(x,0,t)
\nonumber\\
E_d(x,0,t)&=& \kappa_d H_d(x,0,t).
\eea
We should emphasize that this is not intended to be a realistic model
and we only use it to illustrate the typical size of effects that one
might anticipate. 

\begin{figure}
\unitlength1.cm
\begin{picture}(10,13.5)(3.7,2.7)
\includegraphics{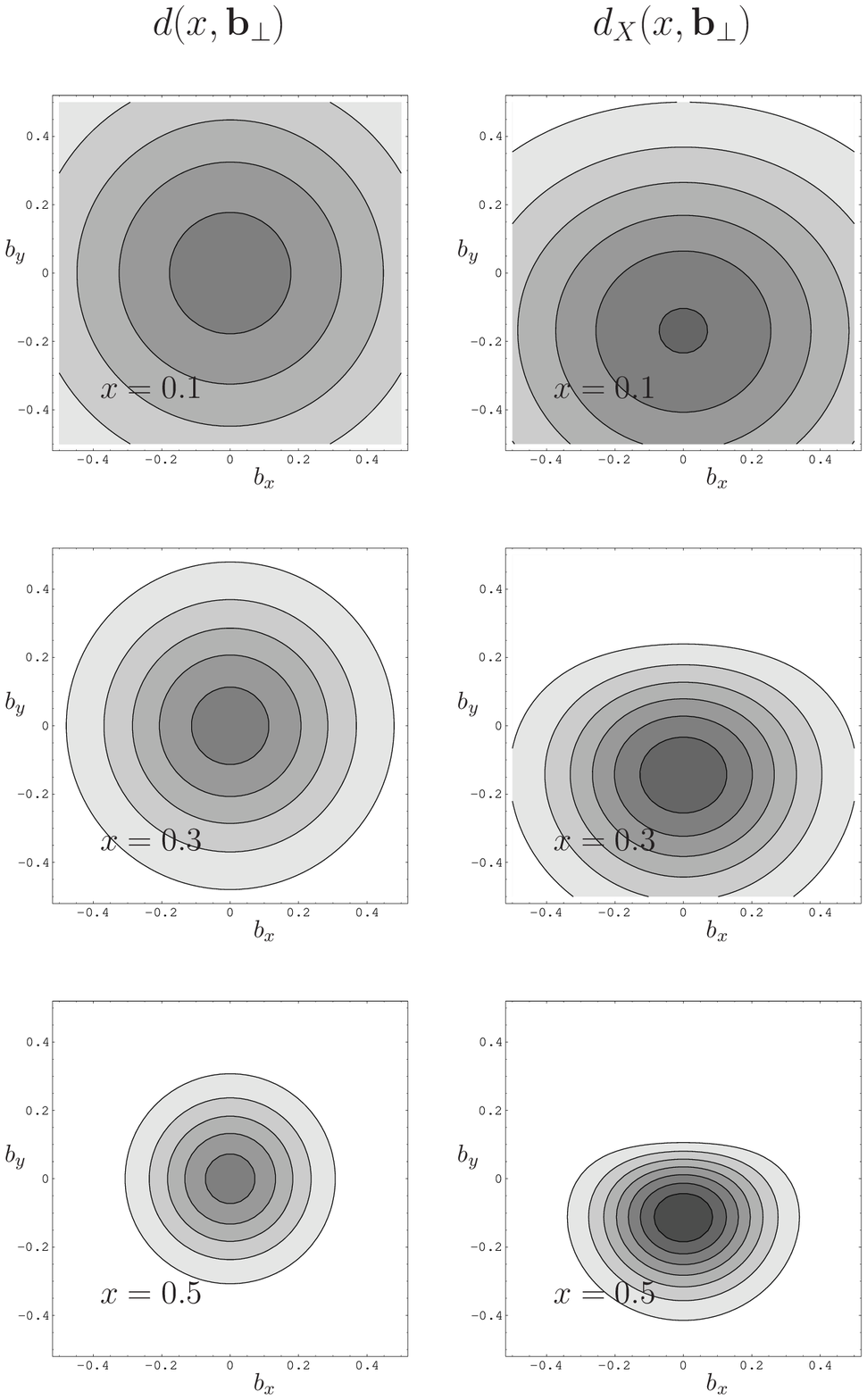}
\end{picture}
\caption{Same as Fig. \ref{fig:panelu}, but for
$d$ quarks. }
\label{fig:paneld}
\end{figure}

The resulting parton distributions in impact parameter space for $u$
and $d$ quarks are shown in Figs. \ref{fig:panelu} and 
\ref{fig:paneld} respectively.
Note that PDFs as well as GPDs decrease significantly 
from $x=0.1$ to $x=0.5$. In order to be able to plot the impact 
parameter dependence we normalized the distributions for each value of
$x$ and both $u$ and $d$ quark distributions to the value of the
longitudinally polarized distribution at $\bp=0$.

The `tomographic slices', i.e. the impact parameter dependences for
a few fixed values of $x$, that are shown in Figs. \ref{fig:panelu} 
and \ref{fig:paneld} clearly demonstrate what should have been clear 
already from our model-independent result above (\ref{eq:dy}):
at larger values of $x$, the $u$ and $d$ quark distributions in a
transversely polarized proton are shifted to opposite sides and
the magnitude of the distortion is such that there is a significant
lack of overlap between the two. Other models for $E(x,0,t)$
\cite{E} yield very similar result since the overall magnitude of
the effect is constrained by the model independent
relation Eq. (\ref{eq:dy}).

Such a large separation between quarks of different flavor, which
is both perpendicular to the momentum and spin of the proton
must have some observable effects. For example, 
in semi-inclusive
photo-production of pions off transversely polarized
nucleons, the $u$ quarks are knocked out predominantly
on one side of the nucleon.
Therefore the final state interaction will be different for pions
produced going to the right compared to those going to the left,
which in turn may lead to a transverse asymmetry of produced pions.
Other examples are flavor
exchange reactions and for given transverse polarization, the
added quarks might be picked up predominantly one one particular side
of the hadron, suggesting a transverse asymmetry of the hadron
production relative to the nucleon spin.
In the next section, we will present a simple model for these final 
state interactions, which together with the transverse asymmetries in the position space distribution of partons, leads to predictions for 
the signs of the transverse asymmetries 
in various hadron production reactions.

\section{Single Transverse Spin Asymmetries}
Many semi-inclusive hadron production experiments show surprisingly 
large transverse polarizations or asymmetries \cite{lambda}.
Moreover, the signs of these polarizations are usually
not dependent on the energy.
This very stable polarization pattern suggests that
there is a simply mechanism that underlies these polarization
effects. In the following, an attempt is made to link
the large transverse distortions of parton distributions
in impact parameter space for transversely polarized nucleons
(baryons) with these transverse single spin asymmetries. 

We will make the following model assumptions for flavor transitions
in high energy scattering events:
In a flavor changing process, as many quarks as possible
(hereafter referred to as ``spectators'') 
originate from the impacting hadron.
Any additional quarks are produced from the breaking
of a string that connects the spectators with the target right
after the impact. Since this string exerts an attractive force
on the ``spectators'' before it breaks, this picture suggests that
the transverse momentum of the final state hadron will point
in the direction given by the side on which the additional
quarks were produced.

Note that this model implicitly focuses on more peripheral
scattering events for describing the signs of baryon polarizations 
at large $x_F$. 
Although these may not be the only possible
events, we expect that central collisions are less likely to produce
the observed pattern of large and only weakly energy dependent 
polarizations. This is supported for example by the 
observation that the polarization of the produced $\Lambda$ hyperons
is particularly large in diffractive production \cite{boros:diff}.

These simple model assumptions, together with the transverse
distortion of quarks in transversely polarized hadrons provide
an intuitive explanation for the large observed transverse
polarization in inclusive hyperon production as we will
demonstrate in the following. For this purpose, let us consider 
for example a $\Lambda$ that is
produced moving to the left of the incident proton beam.
\begin{figure}
\unitlength1.cm
\begin{picture}(10,9)(1,8)
\includegraphics{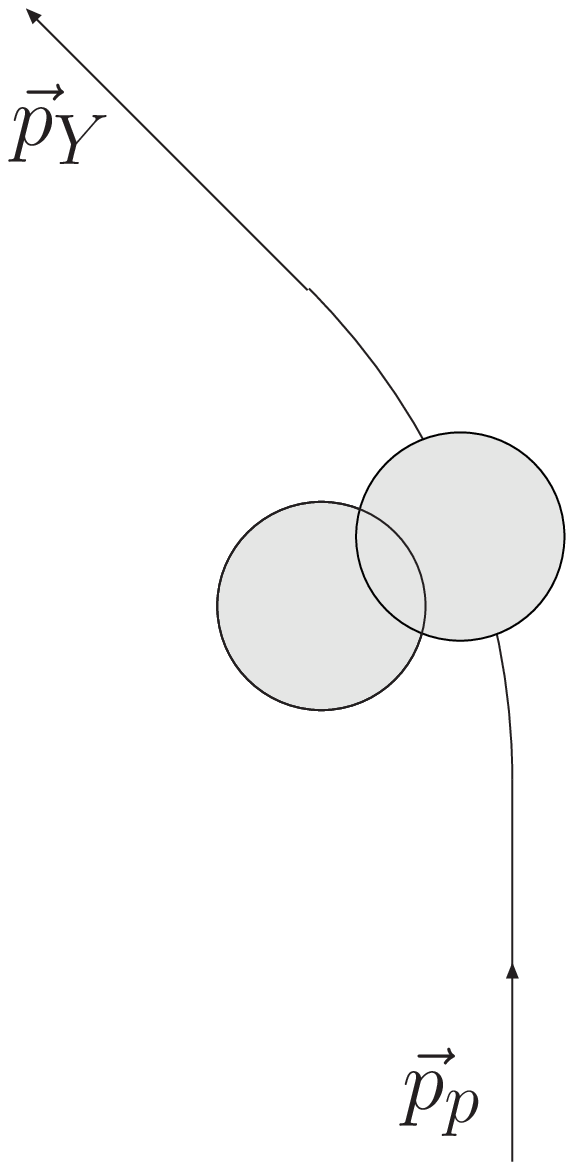}
\end{picture}
\caption{Inclusive $p\longrightarrow Y$ scattering
where the incoming $p$ (from bottom) 
diffractively hits the right side of the target
and is therefore, according to the model assumptions, 
deflected to the left during the reaction.
The $s\bar{s}$ pair is assumed to be produced
roughly in the overlap region, i.e. on the left
`side' of the $Y$.}
\label{fig:collide}
\end{figure}  
Using our model assumptions above, this implies that the $s$ quark
was produced on the left side of the $\Lambda$. Since
$\kappa_{s/\Lambda}>0$, such a state with an $s$-quark produced
on the left side has a much better overlap with a $\Lambda$ that
has spin down (when one looks into the beam direction)
rather than spin up. 
\begin{figure}
\unitlength1.cm
\begin{picture}(10,4)(1,9)
\includegraphics{fshift.ps}
\end{picture}
\caption{Schematic view of the transverse distortion
of the $s$ quark distribution (in grayscale)
in the transverse plane for a transversely
polarized hyperon with $\kappa_s^Y>0$. The view
is (from the rest frame) into the direction of 
motion (i.e. momentum into plane) for a hyperon 
that moves with a large momentum.
In the case of spin down (a), the $s$-quarks get distorted toward
the left, while the distortion is to the right
for the case of spin up (b).}
\label{fig:shift}
\end{figure}  
Therefore, for a $\Lambda$ that has been deflected to the left one 
would expect a polarization that points downward. Following the
usual convention, where the polarization direction
is defined w.r.t. the normal vector
${\vec n} \equiv {\vec p}_{beam}\times {\vec p}_{final}
/|{\vec p}_{beam}\times {\vec p}_{final}|$, the $\Lambda$ should
have negative polarization, which is also what is observed 
experimentally \cite{lambda}.
Likewise, since $\kappa_{s/\Sigma}<0$ and
$\kappa_{s/\Xi}>0$ (Appendix), one would expect that $\Sigma$ and 
$\Xi$ hyperons are produced with polarizations ``up'' and ``down''
respectively when one starts from an incident proton beam and the
hyperon is produced to the left of the beam.

\begin{figure}
\unitlength1.cm
\begin{picture}(10,16)(4.5,11.5)
\includegraphics{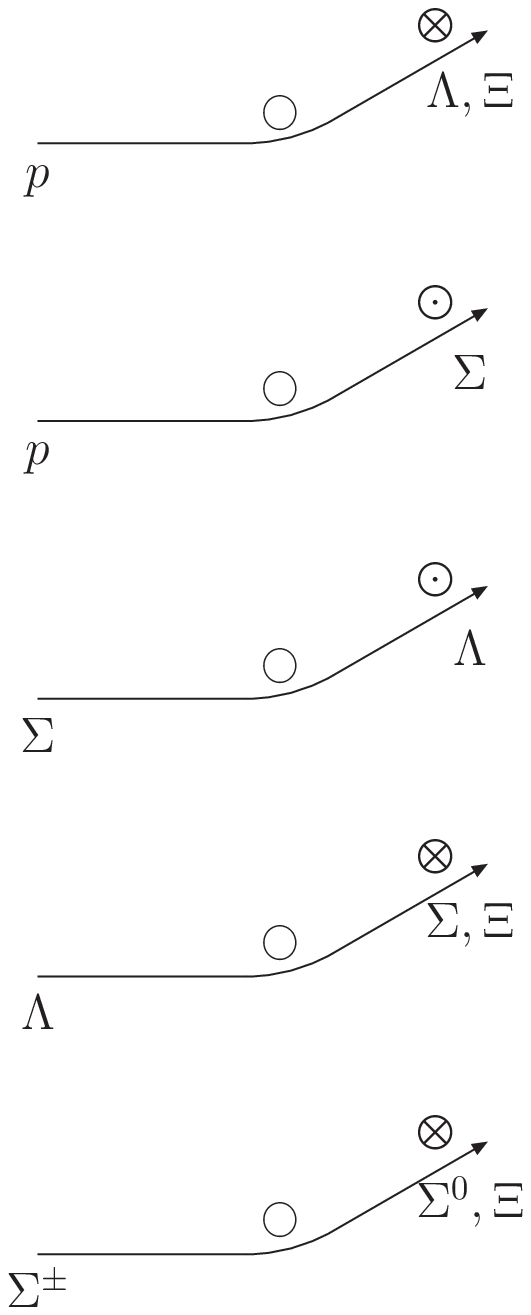}
\end{picture}
\caption{Transverse polarizations of hyperons that are produced from
an unpolarized beam and target (represented by an empty circle). 
According to the model assumptions, the
final state hadron is deflected in the direction given by the
side on which the missing quarks were produced.
$\odot$ and $\otimes$ represent hyperons with spin pointing 
out of the plane and into the plane respectively. 
}
\label{fig:ssa}
\end{figure}

If the incident beam consists of $\Lambda$ or $\Sigma$ hyperons,
then the polarization of produced $\Xi$ hyperons is of course
the same as in the case of incident nucleons since it is still
only $s$ quarks that need to be substituted. However, the
situation changes if one considers $\Lambda \rightarrow \Sigma$
and $\Sigma \rightarrow \Lambda$ production reactions, because
there it is a $u$ or $d$ quark that needs to be substituted.
If we now use that $\kappa_{u/\Lambda}=\kappa_{d/\Lambda}<0 $
and for example $\kappa_{u/\Sigma}>0 $, one finds that the
sign of the polarization of $\Lambda/\Sigma$ produced from a
$\Sigma/\Lambda$ beam is reversed compared to the respective
polarizations that arise when one starts from a nucleon beam
(Fig. \ref{fig:ssa}). However, we should emphasize that
$|\kappa_{u/\Lambda}|$ is only about half as large as
$\kappa_{s/\Lambda}$ and therefore the transverse 
distortion of the $u/d$ quarks in a transversely polarized $\Lambda$
is expected to be smaller than the one of the $s$ quarks.
We therefore
expect that the polarization of $\Lambda$ produced from an
incident $\Sigma$ beam is not only reversed but also significantly
smaller in magnitude than those produced from a proton beam.
For neutron production the spin in the final state is not 
self-analyzing. However, our model also predicts interesting
asymmetries with respect to the spin of the initial state.
\begin{figure}
\unitlength1.cm
\begin{picture}(10,10.3)(4.8,18.)
\includegraphics{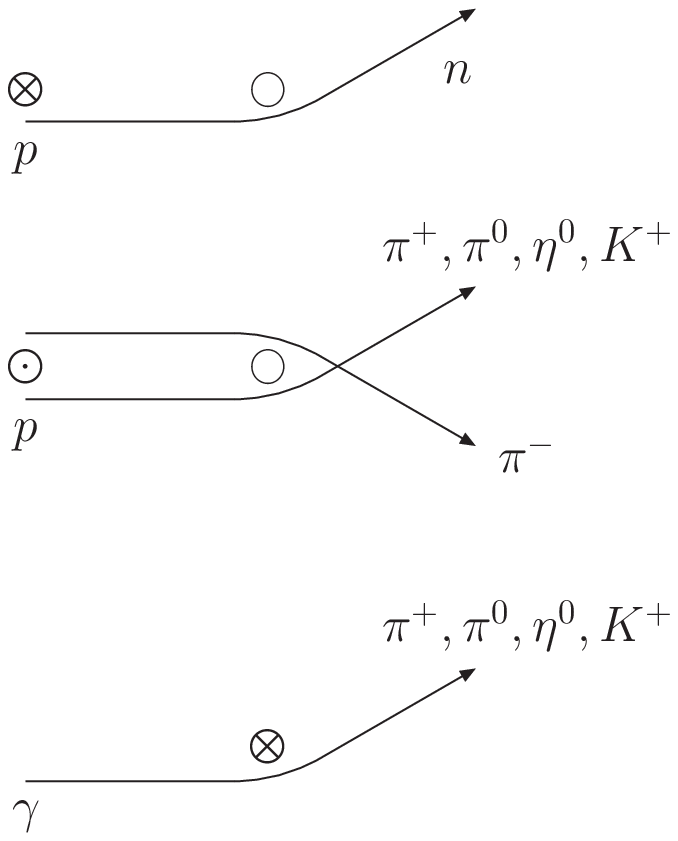}
\end{picture}
\caption{Beam and target spin asymmetries for\\ 
$p \longrightarrow n/meson$ and semi-inclusive
$\gamma \longrightarrow meson$ respectively.
}
\label{fig:ssa2}
\end{figure}  
In order to be converted into a neutron, the proton must strip off
one of its $u$ quarks. A proton that is polarized `down' has its
$u$ quarks shifted to the left of its center of momentum, i.e.
it can strip off a $u$ quark more easily when it passes the target
on the right and, at least within our model, will be more likely to 
result in a neutron that is deflected to the left 
(Fig. \ref{fig:ssa2}). In summary, we therefore expect neutrons to be
more likely to be produced to the left of the beam if the proton spin
is downward and to the right if its spin is upward, corresponding to
a negative analyzing power. This result agrees with a recent 
measurement at RHIC \cite{pton}.

We should emphasize that similar reasoning for inclusive hyperon
production also implies a spin asymmetry with respect to the 
incident proton spin. If we define again a positive analyzing
power $A_N$ if protons with spin up give rise to a final state hadron 
that is deflected to the left, then $p\rightarrow \Lambda$ should 
also have $A_N<0$ since there one also needs to substitute a $u$
quark in the proton. The situation is similar for  $p\rightarrow 
\Xi^{-}$, where both $u$ quarks need to be substituted.
In the case of $p\rightarrow \Sigma^+$, it is the $d$ quark that is
substituted and therefore $A_N>0$.

The beam asymmetries in semi-inclusive meson production can be 
explained similarly. 
In order for a proton to convert into a 
$\pi^+$, one
of its $u$ quarks needs to `go through'. This is most likely to happen
if the $u$ quarks are on the ``far side'' of the interaction zone.
This favors protons with spin up when the proton passes the target on
the right and spin down when it passes on the left side of the target.
If we assume again that the final state interaction that leads to
string breaking is attractive (until the string breaks) then
protons with spin up result in $\pi^+$ that are more likely 
deflected to the right, while protons with spin down are more 
likely resulting in $\pi^+$ that are deflected to the left, i.e.
we expect a positive analyzing power for $p\rightarrow \pi^+$ and
the same for $p\rightarrow K^+$.
For $\pi^-$ we expect a negative analyzing power since there the
leading quark is a $d$ quark, which would be more likely on the
side opposite to the $u$ quarks for a transversely polarized nucleon
and one expects a negative analyzing power.
For $p\rightarrow \pi^0,\eta^0$ the leading quark could be both
$u$ or $d$, but since valence $u$ quarks outnumber the $d$ quarks in 
a proton, one expects that the net analyzing power is again positive, but smaller than for $\pi^+$. These results seem to be consistent with
the pattern that is observed experimentally \cite{ptopi}.

\begin{figure}
\unitlength1.cm
\begin{picture}(10,7)(4.5,18.5)
\includegraphics{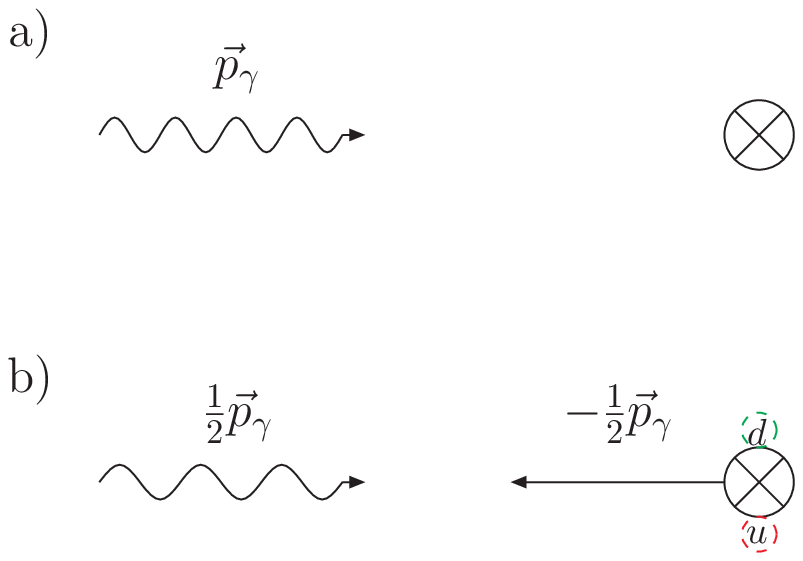}
\end{picture}
\caption{Photon hitting proton target. a) laboratory frame,
b) CM frame. The polarization of the proton is into the plane.
According to the results from Sec. \ref{sec:gpdflip}, the $u$ quarks
(schematically indicated by a dashed circle)
are shifted down.}
\label{fig:ssa3}
\end{figure}

In order to understand target spin asymmetries, it is useful to
analyze the process in the CM frame where the projectile and the
target have initially opposite momenta. As an example, let us
consider the target spin asymmetry in semi-inclusive
electro-production of pions on a 
transversely polarized proton target (Fig.\ref{fig:ssa3}).
For a target polarization that is into the plane, and applying the
results from Sec. \ref{sec:gpdflip}, the $u$ quark 
distribution in the CM frame is shifted down, while the $d$ quark
distribution is shifted up. Semi-inclusive
photo-production of mesons with a
$u$ valence quarks (e.g. $\pi^+,\pi^0,\eta^0,K^+$) occurs
dominantly through photons that initially interact with a $u$ quark 
in the target, which later fragments into the meson. 
Applying again our model assumption from above, i.e. using that the
QCD string deflects the $u$ quark toward the center, we conclude that
the mesons with a valence $u$ quark are produced preferentially
in the up direction (Fig. \ref{fig:ssa2}) within this model, 
i.e. to the left if one looks into the direction of 
the photon momentum and the spin of the proton is down. 
For mesons without valence $u$ quarks, such as the $\pi^-$,
there are two competing effects: when the photon hits the $d$ quark 
first then our argumentation above would favor $\pi$ deflected in the
direction opposite to $\pi^+$, since the $d$ quarks are, for a given
polarization of the proton, shifted in the direction opposite to
the $u$ quarks. However, the contribution from `disfavored' 
fragmentation $u\rightarrow \pi^-$ is enhanced due to the  fact
that the photon is much more likely to hit a $u$ than a $d$ quark in
the proton and therefore the resulting asymmetry is not immediately
obvious.

\section{Discussion}

Our model for generating the polarizations and spin asymmetries is
much too crude to make detailed quantitative predictions about the 
size of the effects. However, the model matches the observed signs and
provides a natural explanation for the fact that the observed effects
are very large. We not only obtain a unified description for
polarization and single spin asymmetry experiments but at the same
time develop a link between these spin observables and parton
distributions in impact parameter space.

There have been a number of models attempting to explain polarizations
observed in hyperon production experiments and it would be beyond the
intended scope of this article to provide a detailed comparison with
all of them \footnote{A nice recent review on the subject can be
found in Ref. \cite{boros}.}, but we would still like to point out
a few similarities and differences.

It is interesting to compare our attempt to link asymmetries of 
parton distributions in impact 
parameter space with single spin asymmetries with attempts to 
link asymmetries of unintegrated parton densities with the
single spin asymmetries \cite{sivers}. 
The main difference between these two approaches is that
we start from a transverse asymmetry in position space.
The final state interaction of the outgoing quark the converts
the position space asymmetry into an asymmetry for the 
transverse momentum of the final state hadron.
The Sivers effect is complementary in that it starts already from
an asymmetry in the transverse momenta of the unintegrated parton 
densities. Of course, since 
${\vec S}\cdot ({\vec k}_q\times {\vec p})$ is not a Lorentz scalar
under time-reversal, such an asymmetry in the unintegrated parton
densities appears only if the final state interaction is included
into their definition via appropriate Wilson lines \cite{collins2},
i.e. in a sense the final state interactions are already included
in the definition of these unintegrated parton densities.
From that point of view, these two approaches are complementary
attempts to explain single transverse spin asymmetries, which both
have in common that they rely on final state interactions, although
the technical details are very different and it remains to be 
seen whether the Sivers model and this work
describe the same physics but only
from different angles or whether they actually describe different
physical mechanisms.
 
The pattern of signs that we predict resembles very much that of other
semi-classical models. This should not come as a surprise since the
orbital angular momentum of quarks plays an important role in many
of these models. In our model the connection
with quark orbital angular momentum appears because the same 
GPD that describes the transverse 
distortion of PDFs in impact parameter space [namely $E_q(x,0,\Dps)$]
also appears in a sum-rule for the angular momentum carried by the
quarks \cite{ji}. Nevertheless there are few differences to these 
models. For example, in a model where the interaction is assumed to
happen at the front of the hadron, the left-right asymmetries are 
generated by the transverse momentum of quarks with orbital angular 
momentum at the front side \cite{surface}. Such a model would in 
general predict
exactly the same polarization/asymmetry pattern as our model, with
the exception of reactions where the incoming projectile is a photon.
In that case the absorption is weak and it is not legitimate to argue 
that the interaction of the photon with the target should be a 
surface effect. Therefore, models where the polarization results
as a combination between the initial state interaction and the quark
orbital angular momentum would only predict a very small transverse
single spin asymmetry in semi-inclusive 
photo-production experiments. In our model,
the impact parameter space asymmetry is translated into a
momentum asymmetry of the outgoing hadron as a result of the final 
state interaction and therefore the expected asymmetries in 
semi-inclusive photo-production experiments are of the same order of 
magnitude as in hadro-production experiments.

Like in Ref. \cite{collins}, the physical mechanism that eventually
leads to polarization/asymmetries in our model is the final
state interaction of the fragmenting quark(s). It would be 
interesting to see if the similarity between these two mechanisms 
goes beyond this simple observation. 

It is conceivable that studying spin transfers, i.e. the 
correlation $D_{NN}$ between the  transverse polarization 
of the produced baryon 
and the transverse polarization of the beam, leads to further
insights about the mechanism for transverse polarizations because
it may help to differentiate between various models.
In our model a correlation between the spins of the initial and final baryon arises because the transverse distortion of impact parameter
dependent PDFs in transversely polarized hadrons leads to both
polarizations as well as transverse single spin asymmetries.
The correlation between the initial and final state transverse spin
is such that the removed valence quark should be on the same side
of the initial state baryon as the substituted valence quark in
the final state baryon.
Therefore the sign of $D_{NN}$ is determined by the
sign of the product of the $\kappa_q$ for the valence quark that 
stripped of and the quark that is substituted for it.
For example, in the $p\rightarrow\Lambda$ transition, a $u$ quark 
needs to be substituted by an $s$ quark. Since 
$\kappa_{u/p}*\kappa_{s/\Lambda}>0$ we would expect a positive
spin transfer in this case. 

\section{Summary}

Generalized parton distributions for purely transverse
momentum transfer can be related to the distribution of partons
in the transverse plane. When the nucleon is polarized in the
transverse direction (e.g. transverse w.r.t. its momentum in the
infinite momentum frame) then the distribution of partons in the 
transverse plane is no longer axially symmetric. The direction
of the transverse distortion is perpendicular to both the spin and
the momentum of the nucleon. Classically the effect can be understood
as a superposition of the translatory motion of the partons along
the momentum of the nucleon with the orbital angular motion of 
partons in the nucleon. The sign and magnitude of the distortion
of (unpolarized) PDF in impact parameter space can be expressed 
in terms of the helicity-flip generalized parton distribution
$E_q(x,0,-\Dps)$. Since $\int dx E_q$ can be related to the Pauli form
factor $F_{2,q}$ for flavor $q$, one can thus relate the resulting
transverse flavor dipole moment of the distorted parton distributions
to the anomalous flavor-magnetic moment $\kappa_{q/p}$ in the proton.
We are thus able to link the transverse distortion of partons to
the magnetic properties of the nucleon which leads to a 
model-independent prediction for the resulting transverse flavor
dipole moments that are on the order of $0.1-0.2 \,$ fm.

Such a large transverse dipole polarization for quarks of different
flavor should also have observable effects in semi-inclusive
hadron production experiments. We introduced a simple model to 
translate the transverse asymmetry of the parton distributions in
impact parameter space into transverse asymmetries of the
produced hadrons. The basic idea of the model is that the leading 
quark(s),\footnote{In photo-production experiments, 
the `leading quark' in the model 
is simply the struck quark, while in hadro-production experiments
the `leading quarks' are spectator quarks from the incident hadron.}
before they fragment into the observed hadron, experience
an attractive force from the QCD string before the string breaks.
This attractive force between the produced outgoing hadron and the
target remnant leads to the left-right asymmetry in the observed 
hadron distributions.

We use this model to explain or predict a number of 
$baryon \longrightarrow baryon^\prime$ experiments, where the
transverse distortion of transversely polarized baryons favors
certain final polarization states and therefore leads to 
transversely polarized baryons in the final state.
We argue that the large transverse hyperon polarization at high 
energies that is
observed in these experiments is naturally explained due to the fact 
that the transverse flavor dipole moment of transversely polarized
baryons in the infinite momentum frame is also very large.
A similar mechanism is used to explain the asymmetry 
in semi-inclusive meson production using either a 
transversely polarized
proton beam or incident virtual photons hitting a transversely
polarized target.

Acknowledgments: I appreciate several interesting discussions with
G. Bunce and N. Makins and comments from G. Schnell. 
This work was supported by a grant from DOE (FG03-95ER40965). 
\appendix
\section{$SU(3)$ analysis of baryon magnetic moments}
\label{app}
We use a notation, where $F_2^{q/B}$ denotes the the
Pauli form factor $F_2$ defined as the
matrix element of a vector current with flavor $q$,
i.e. $\bar{q}\gamma^\mu q$ between states of the baryon $B$.
It is related to the usual electromagnetic form factor for
that baryons using
\be
F_2^B(Q^2) = \frac{2}{3}F_2^{u/B}(Q^2)-\frac{1}{3}F_2^{d/B}(Q^2)-
\frac{1}{3}F_2^{s/B}(Q^2).
\ee
For the transverse flavor dipole moments, we need to know the
the anomalous magnetic moment contributions for each quark flavor
and each baryon
\be
\kappa_{q/B}\equiv F_2^{q/B}(0).
\ee
Experimentally, little is known beyond the electro-magnetic linear
combination $\sum_qe_q\kappa_{q/B}$ for a few baryons. For our 
purposes, namely explaining the signs of various asymmetries, 
it will be sufficient to know the sign and order of magnitude of
the the $\kappa_{q/B}$. Therefore, we will use
$SU(3)$-flavor symmetry which should be sufficient for an accuracy
of a couple of $10\%$ to estimate the $\kappa_{q/B}$. The only input
that we use are the anomalous magnetic moments of the proton and neutron
\be
\kappa^p&=&\frac{2}{3}\kappa_{u/p} -\frac{1}{3}\kappa_{d/p}
-\frac{1}{3}\kappa_{s/p}=1.79\nonumber\\
\kappa^n&=&\frac{2}{3}\kappa_{u/n} -\frac{1}{3}
\kappa_{d/n} -\frac{1}{3}\kappa_{s/n}= -1.91
\ee
and we will assume that $\kappa_{s/p}\approx 0$.\footnote{Although
$\kappa_{s/p}$ is not known very accurately, it is nevertheless 
clear that its numerical value is significantly smaller than
$\kappa_{u/p}$ and $\kappa_{d/p}$ and it should therefore be 
justified to neglect its contribution for the kind of estimate that 
we are interested in.}
Using isospin symmetry, this implies
\be
\kappa_{u/p} &=& 2\kappa_p +\kappa_n +\kappa_{s/p} \approx 1.67
\nonumber\\
\kappa_{d/p} &=& 2\kappa_n +\kappa_p +\kappa_{s/p} \approx -2.03 .
\label{eq:kappaud}
\ee
If one assumes $SU(3)$ symmetry, then the flavor magnetic moments 
for baryons of type $aab$ are trivially related to the ones in the
proton, using $\kappa_{a/B}=\kappa_{u/p}$,
$\kappa_{b/B}=\kappa_{d/p}$, and $\kappa_{c/B}=\kappa_{s/p}$, which
implies for example
\be
\kappa_{s/\Sigma} &=& \kappa_{d/p} \approx -2.03 
\nonumber\\
\kappa_{s/\Xi} &=& \kappa_{u/p} \approx 1.67 .
\ee
The $\Lambda$ is less trivial, but a straightforward $SU(3)$ analysis
yields
\be
\kappa_{s/\Lambda}=\frac{2}{3}\kappa_{u/p} -\frac{1}{3}\kappa_{d/p}
+\frac{2}{3}\kappa_{s/p} \approx 1.79.
\ee
For flavor changing transitions among hyperons, we also need the
$u/d$ moments
\be
\kappa_{u/\Sigma^+} &=& \kappa_{u/p} \approx 1.67\nonumber\\
\kappa_{u/\Lambda} = \kappa_{d/\Lambda} &=& \frac{1}{6}
\kappa_{u/p} + \frac{2}{3}\kappa_{d/p} + \kappa_{s/p}
\approx -0.98.
\ee

\end{document}